# 3D-CLMI: A Motor Imagery EEG Classification Model via Fusion of 3D-CNN and LSTM with Attention

Shiwei Cheng [a,b*], Yuejiang Hao [a]

[a] School of Computer Science, Zhejiang University of Technology, Hangzhou 310023, China

[b] School of Design, Shanghai Jiao Tong University, Shanghai 200240, China

* Corresponding author. E-mail address: 249401866@qq.com

**Abstract**

Due to the limitations in the accuracy and robustness of current electroencephalogram (EEG) classification algorithms, applying motor imagery (MI) for practical Brain-Computer Interface (BCI) applications remains challenging. This paper proposed a model that combined a three-dimensional convolutional neural network (CNN) with a long short-term memory (LSTM) network with attention to classify MI-EEG signals. This model combined MI-EEG signals from different channels into three-dimensional features and extracted spatial features through convolution operations with multiple three-dimensional convolutional kernels of different scales. At the same time, to ensure the integrity of the extracted MI-EEG signal temporal features, the LSTM network was directly trained on the preprocessed raw signal. Finally, the features obtained from these two networks were combined and used for classification. Experimental results showed that this model achieved a classification accuracy of 92.7% and an F1-score of 0.91 on the public dataset BCI Competition IV dataset 2a, which were both higher than the state-of-the-art models in the field of MI tasks. Additionally, 12 participants were invited to complete a four-class MI task in our lab, and experiments on the collected dataset showed that the 3D-CLMI model also maintained the highest classification accuracy and F1-score. The model greatly improved the classification accuracy of users' motor imagery intentions, giving brain-computer interfaces better application prospects in emerging fields such as autonomous vehicles and medical rehabilitation.

## 1. Introduction

Motor imagery (MI), as one of the commonly used paradigms in brain-computer interfaces (BCIs), involves signals generated in the sensory-motor cortex area of the brain. These signals reflect the brain activity when a person imagines the movement of a particular part of the body. By decoding these signals, corresponding electroencephalogram (EEG) data can be converted into control commands and sent to external devices for manipulation, such as brain-controlled robots[1], exoskeletons[2], and autonomous vehicles[3]. Therefore, improving the classification accuracy of motor imagery EEG signals (MI-EEG) is of significant importance.

However, MI-EEG signals are highly stochastic and non-stationary, making them challenging to work with. The presence of weak signals amid a large amount of noisy data results in a very low signal-to-noise ratio, making feature extraction and classification of MI-EEG signals a highly challenging task.

To enhance the accuracy of MI-EEG signal classification, it is necessary to perform feature extraction tailored to three critical characteristics of EEG signals, namely, the local correlation of EEG signals, the correlation between temporal sequence, and the inter-channel correlation. EEG signals exhibit local correlation in spatial terms, meaning that signals from adjacent spatial regions are highly correlated. In recent years, convolutional neural network (CNN) algorithms[4] have been robustly applied to data processing related to brain-computer interfaces. They not only extract correlated features from neighboring spatial regions of EEG signals but also extract the correlation features of EEG signals between adjacent EEG channels at the same time. Furthermore, given the strong inter-channel correlations in EEG signals, 3D convolutional kernels are utilized for feature extraction, replacing the commonly used 2D convolutional kernels in MI-EEG signal classification. This approach treats EEG channels as an additional dimension for extracting local features. Nevertheless, due to the strong temporal dynamics of EEG signals, CNN often struggles to extract their temporal features. In such cases, the recurrent neural network (RNN) is typically employed to extract temporal features. However, the traditional RNN can encounter issues like gradient explosion or vanishing gradients during training, making them unable to effectively utilize early-stage information. To address these problems, the long short-term memory network (LSTM) is introduced. LSTM mitigates gradient explosion and vanishing gradients by introducing memory units and preserving historical information[5]. Additionally, the incorporation of attention mechanisms allows the model to allocate distinct attention weights to different parts of the input data, enabling it to focus more on the most relevant portions. This aids in capturing contextual information and enhancing the model's performance.

This paper contributes to the field by employing a fusion of 3D CNN and LSTM with attention for feature extraction from MI-EEG signals and subsequently classifying MI-EEG signals (hereinafter referred to as 3D-CLMI). Through the use of the public dataset and a four-class motor imagery task, this study demonstrates that the model can accurately extract features from EEG signals in motor imagery tasks, thereby enhancing the classification prediction performance.

## 2. Related Work

### 2.1. Machine Learning Algorithms

In brain-computer interfaces, traditional machine learning algorithms primarily include methods such as common spatial patterns (CSP), power spectral density (PSD), and wavelet transform (WT). These methods aim to address the challenge of low signal-to-noise ratio in EEG signals. CSP operates on the principle of matrix diagonalization to select an optimal set of spatial filters and project data, maximizing the variance differences between different classes. This yields highly discriminative feature vectors. Within CSP, the filter bank common spatial pattern (FBCSP)[6] method had shown superiority over other approaches. In FBCSP, the filter bank typically comprised a series of bandpass filters, each responsible for extracting signals within specific frequency ranges. CSP was then applied to extract features from each filtered sub-band. PSD is a frequency spectrum analysis method used to describe the energy distribution of signals across different frequency components[7]. In EEG signal analysis, PSD methods show the activity of EEG signals within various frequency ranges, unveiling characteristic patterns of brain activity. Ng et al. had employed PSD-based Welch and Burg methods for feature extraction from EEG signals[8]. Among them, the Burg method, which minimizes forward and backward prediction errors to satisfy the Levinson-Durbin recursion[9], demonstrated superior feature extraction performance. Wavelet transform (WT) is a transformative analysis method that inherits the idea of localizing short-time Fourier transforms and overcomes the limitation of fixed window sizes by providing a time-frequency window that adapts to frequency changes. It is used for time-frequency analysis and feature extraction from EEG signals. Kant et al. proposed a method that combines continuous wavelet transform (CWT) with transfer learning to address motor imagery classification tasks[10]. By applying CWT with different scales of wavelet functions to analyze EEG signals, they extracted EEG signal features from various frequency ranges for motor imagery classification. However, traditional machine learning methods primarily focus on extracting energy and time-frequency features from EEG signals, often overlooking the temporal and spatial features of EEG data. This limitation results in incomplete feature extraction, leading to insufficient classification accuracy in motor imagery tasks.

### 2.2. Deep Learning Algorithms

Traditional machine learning methods like CSP struggle to effectively extract spatial features from EEG signals. The convolutional neural network, on the other hand, can learn robust spatial features for EEG signal classification[11]. Consequently, in recent years, CNN has been widely applied to the problem of MI-EEG signal classification. One of the advantages of CNN is its ability to automatically extract features from EEG signals without the need for manually defined feature extraction rules. This eliminates the limitations of manual feature design and allows CNN to exhibit better generalization capabilities across different tasks and datasets[12-13].

Mattioli et al. introduced a 1D CNN model applied to MI-EEG signal classification[14]. In their proposed 1D CNN model, they employed a 10-layer one-dimensional convolutional neural network for MI-EEG signal classification after data augmentation, achieving good classification results. While 1D convolutions have shown promise, 2D convolutions can better account for the spatial structure of EEG signals and extract spatial features. Ieracitano et al. proposed a 2D CNN model for motor imagery classification[15]. In their model, they used power spectral density for feature extraction and employed a 2D CNN network for training and feature classification. They achieved an 83.3% classification accuracy on a three-class motor imagery dataset. Tang et al. introduced a multi-scale hybrid network called MSHCNN for motor imagery classification tasks [16]. MSHCNN consisted of a 1D CNN block and a 2D CNN block, each containing multiple convolutional kernels of varying sizes. This network can classify EEG signals with minimal preprocessing and achieved an 85.2% classification accuracy. The above models used 1D convolution kernels and 2D convolution kernels to convolution MI-EEG signals to extract spatial features. However, the models mentioned above did not fully consider that MI-EEG signals were continuous signals with strong temporal characteristics. Solely using convolutional neural networks may not effectively extract temporal features. Additionally, these models applied convolution to EEG images from individual EEG channels and treated EEG channels as feature channels, rather than considering EEG channels as a convolutional dimension for feature extraction. This oversight neglected the local correlations between EEG channels.

Due to the temporal features of MI-EEG signals, purely applying convolutional operations may not fully extract the temporal features present in the raw data. In EEG signal classification, it is common to incorporate long short-term memory networks to extract temporal features. LSTM addresses the issues of gradient vanishing and exploding gradients in traditional RNN by introducing gate mechanisms. Its core idea involves controlling the flow of information through structures known as "gate", including input gate, forget gate, and output gate[17]. Wang et al. proposed a brain EEG classification algorithm based on a CNN-LSTM feature fusion network[18]. In this model, CNN and LSTM networks were connected serially, and the raw EEG signal was first convolved, and then the convoluted signal was handed over to the LSTM network to extract spatial and temporal features. Khademi et al. also introduced a CNN-LSTM fusion approach[19]. They incorporated transfer learning into their network, leveraging classic networks like ResNet-50 and Inception-v3 for feature extraction, and then these features were concatenated with an LSTM network to address the issue of their own shallower convolutional neural network and incomplete feature extraction, achieving a classification accuracy of 87.6%.

The two mentioned models have effectively considered the temporal features of MI-EEG signals. However, the drawbacks of this serial CNN-LSTM network model were also evident: during the process of applying convolution to EEG signals and passing them as input to the LSTM network, some temporal features may be lost due to the convolution operation. This made it challenging for the LSTM to fully extract the temporal features of EEG signals. Additionally, these two models did not address the issue of weighting different parts of the data during temporal feature extraction. This oversight could potentially lead to the model learning excessively from less relevant data features, resulting in suboptimal classification performance.

The existing models that use 3D CNN for motor imagery classification have not taken into account the local correlation among EEG channels and have not employed corresponding convolution strategies. In contrast, the LSTM network with attention considers the strong temporal feature of MI-EEG signals, enabling effective extraction of signal temporal features. Moreover, the attention mechanism can enhance the model's focus on different parts of the input sequence[20], allowing the model to concentrate on segments relevant to the current task and better handle long-distance dependencies when dealing with long sequences[21]. Additionally, the parallel network structure of CNN and LSTM enables LSTM to extract complete temporal features. Therefore, this paper combines 3D CNN with LSTM with attention mechanisms, effectively balancing the spatial features of MI-EEG signals and the local correlation between EEG channels, resulting in the extraction of more comprehensive and complete features, thus improving the accuracy of motor imagery classification tasks.

## 3. 3D-CLMI

This paper introduces a parallel network structure (3D-CLMI) composed of 3D CNN and LSTM for classifying EEG data in motor imagery tasks. The structure is illustrated in Figure 1, where both the 3D CNN and LSTM modules work in parallel to extract features, which are then combined for classification. The main workflow is as follows:

(1) Preprocessed EEG data is transformed into a two-dimensional matrix and different channel EEG data is stacked and combined to create three-dimensional EEG data. Multiple 3D convolution kernels of varying sizes are used to extract spatial features from the EEG data, resulting in EEG data I.

(2) Preprocessed EEG data is directly processed using LSTM with attention mechanisms to extract temporal features, resulting in EEG data II.

(3) The two sets of EEG data are expanded into one dimension and merged through the flatten layer, and the merged EEG data is put into the fully connected layer for classification.

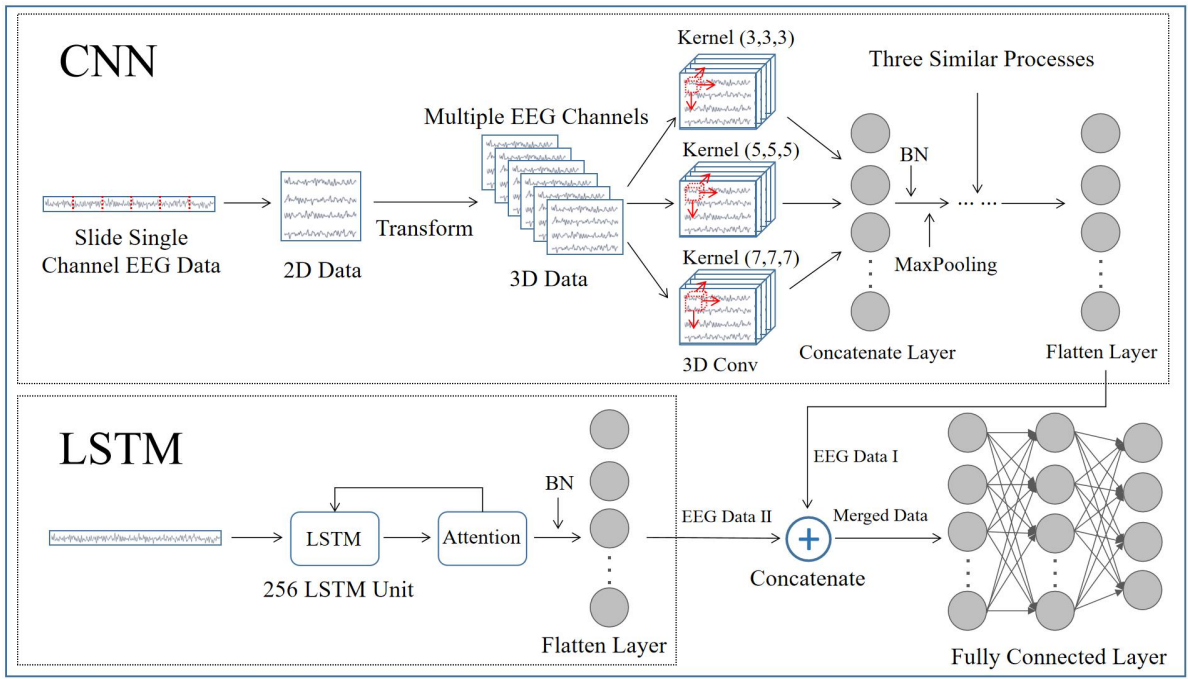

Figure 1　3D-CLMI structure diagram

There are three main preprocessing steps for EEG data, namely spatial filtering, frequency filtering, and data augmentation. Considering the high non-stationarity and low signal-to-noise ratio of EEG signals, we adopt the common average reference (CAR) method for spatial filtering. Since the motor imagery signal generated by the cerebral cortex is in a special frequency band, frequency filtering is used to extract the EEG signals required for motor imagery classification. Data augmentation increases the number of training samples to improve the training quality of the model and obtain better accuracy. The above three steps will be detailed introduced in section 4.2 of this paper.

In our model, the CNN module consists of 4 convolutional layers, with each layer employing 3 convolution kernels of different sizes. We believe that using convolution kernels of different sizes for convolution can effectively extract features of different granularities. After each convolution layer, there is a batch normalization (BN) layer and a max-pooling layer. The BN layer is used for data normalization, while the max-pooling layer helps improve computational efficiency. Similar convolution, batch normalization, and max-pooling operations will be repeated three times to ensure the extraction of deeper feature information. As for the LSTM module, it comprises an LSTM network with 250 LSTM units and an attention layer for parameter weight adjustment. By adding an attention layer, the model can better judge the focus of the temporal features of the motor imagery task and obtain better feature extraction results. The above two modules work in parallel to perform feature extraction to obtain better feature extraction results and improve the classification accuracy of motor imagination tasks. The specific parameter settings for each layer in the 3D-CLMI model are detailed in Table 1.

Table 1  Parameters of 3D-CLMI model.

| | Layer | Filters | Filling | Kernel | Output | Activation |
|---|---|---|---|---|---|---|
| | Input | | | | (None, 30, 30, 22, 1) | |
| | Conv3d_1a | 32 | same | (3,3,3) | (None, 30, 30, 22, 32) | ReLU |
| | Conv3d_1b | 32 | same | (5,5,5) | (None, 30, 30, 22, 32) | ReLU |
| | Conv3d_1c | 32 | same | (7,7,7) | (None, 30, 30, 22, 32) | ReLU |
| | Conv3d_1 | | | | (None, 30, 30, 22, 96) | |
| | BN Layer_1 | | | | | |
| | MaxPooling_1 | | same | (2,2,2) | (None, 15, 15, 11, 96) | ReLU |
| | Conv3d_2a | 64 | same | (3,3,3) | (None, 15, 15, 11, 64) | ReLU |
| | Conv3d_2b | 64 | same | (5,5,5) | (None, 15, 15, 11, 64) | ReLU |
| | Conv3d_2c | 64 | same | (7,7,7) | (None, 15, 15, 11, 64) | ReLU |
| | Conv3d_2 | | | | (None, 15, 15, 11, 192) | |
| | BN Layer_2 | | | | | |
| CNN | MaxPooling_2 | | same | (2,2,2) | (None, 8, 8, 6, 192) | ReLU |
| | Conv3d_3a | 128 | same | (3,3,3) | (None, 8, 8, 6, 128) | ReLU |
| | Conv3d_3b | 128 | same | (5,5,5) | (None, 8, 8, 6, 128) | ReLU |
| | Conv3d_3c | 128 | same | (7,7,7) | (None, 8, 8, 6, 128) | ReLU |
| | Conv3d_3 | | | | (None, 8, 8, 6, 384) | |
| | BN Layer_3 | | | | | |
| | MaxPooling_3 | | same | (2,2,2) | (None, 4, 4, 3, 384) | ReLU |
| | Conv3d_4a | 128 | same | (3,3,3) | (None, 4, 4, 3, 128) | ReLU |
| | Conv3d_4b | 128 | same | (5,5,5) | (None, 4, 4, 3, 128) | ReLU |
| | Conv3d_4c | 128 | same | (7,7,7) | (None, 4, 4, 3, 128) | ReLU |
| | Conv3d_4 | | | | (None, 8, 8, 6, 384) | |
| | BN Layer_4 | | | | (None, 4, 4, 3, 384) | |
| | MaxPooling_4 | | same | (2,2,2) | (None, 2, 2, 2, 384) | ReLU |
| | Flatten Layer_1 | | | | (None, 3072) | |
| | Input | | | | (None, 30, 30, 22, 1) | |
| | Reshape Layer | | | | (None, 900, 22) | |
| LSTM | LSTM | 256 | | | (None, 256) | |
| | Attention | | | | (None, 256) | |
| | BN Layer_5 | | | | | |
| | Flatten Layer_2 | | | | (None, 256) | |
| | Concatenate | | | | (None, 3328) | |
| FC | Dropout | | | | p = 0.3 | |
| | Fully Connected Layer | 4 | | | (None, 4) | softmax |

## 4. Dataset and Preprocessing
### 4.1. Dataset
The validation was conducted on the public dataset BCI Competition IV 2a[22]. Additionally, 12 participants were invited to take part in the motor imagery experiments, and their EEG signals were collected for training and prediction. Next, the data acquisition process for each of the two datasets will be described.

#### 4.1.1. BCI Competition IV 2a
The BCI Competition IV 2a dataset consisted of EEG data from 9 participants. This dataset was derived from a four-class motor imagery task involving left-hand, right-hand, feet, and tongue movements. Each participant conducted two sets of experiments on different days, with the data from each set used for training and testing the classifier. For each set, it can be further divided into 6 runs, each containing 48 trials. Before the experiment, the participants were seated comfortably in a chair in front of a computer screen. At the beginning of each trial ($t$=0s), a fixed crosshair appeared on the black screen, accompanied by a brief auditory cue. Two seconds later ($t$=2s), an arrow pointing left, right, down, or up (corresponding to the four classes: left-hand, right-hand, feet, and tongue movements) appeared on the screen for approximately 1.25 seconds, instructing the participant to imagine the corresponding action. Each participant needed to perform motor imagery until the crosshair disappeared from the screen ($t$=6s). At the end of each trial, the participants can take a short break until the screen turns black again. The average duration of each trial was approximately 8 seconds, as illustrated in Figure 2.

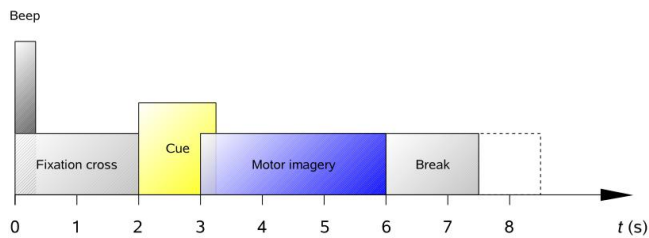

Figure 2    Data acquisition paradigm[22].

#### 4.1.2. Experimental Dataset
In addition to the aforementioned public dataset, an experimental setup was designed to collect MI-EEG signals from users for a four-class motor imagery task involving left hand, right hand, feet, and tongue movements. This setup was used to validate the proposed model in this paper.

12 participants were recruited for the study, consisting of 6 females and 6 males, with an average age of 24 years. Additionally, all participants passed the Ishihara color blindness test[23], had normal or corrected-to-normal vision, and were right-handed as confirmed by the Edinburgh Handedness Inventory[24]. All experiments were conducted with approval from the ethics committee of the author's university. Before the experiments, participants were informed about the experimental procedures and provided their informed consent.

A 64-channel EEG cap was used for data acquisition, with a sampling frequency of 1000 Hz. Before the start of the experiment, experimenters placed the EEG cap on the participants. Following the international 10-20 system[25], the 64 EEG channels and 2 reference channels (REF and GND) on the EEG cap were carefully positioned to correspond to specific locations on the participant's scalp. Electrode gel was applied to each electrode, and real-time impedance monitoring was conducted to ensure minimal resistance between the electrodes and the scalp. Before commencing the experiments, participants were instructed to sit in front of the screen with their eyes approximately 50 centimeters away from the screen and maintained a stable posture. The experimental scene was illustrated in Figure 3. The 64 channels of the EEG cap were displayed in Figure 4. Following references from[26], we selected five channels highly relevant to motor imagery rhythms for data collection in the dataset, namely C3, C4, F3, F4, and Cz channels (highlighted with blue circles).

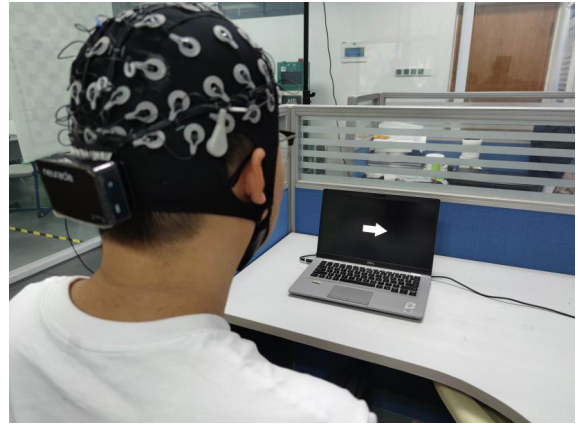

Figure 3    The participants performed motor imagination data collection.

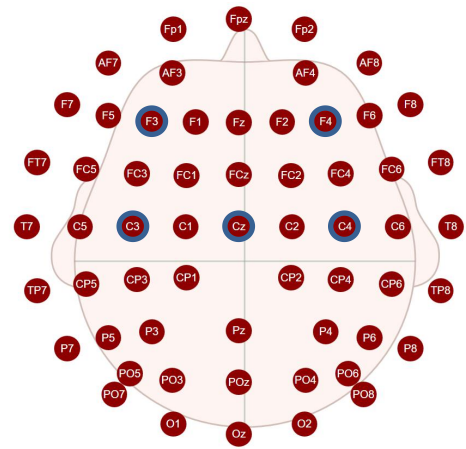

Figure 4    Location of EEG channel distribution.

The experimental paradigm of this experiment was shown in

Figure 5. The duration of each trial was 9 seconds: the first second (*t*=0) was the preparation time, during which a white cross will be displayed on the screen to prompt the subject to prepare. Subsequently (*t*=1), left, right, down, or up arrows will appear on the screen, corresponding to the motor imagination task of the left hand, right hand, feet, or tongue movement respectively, lasting for 1 second. After one second (*t*=2), the participant needed to imagine the corresponding body movements according to the direction of the arrow on the screen in the next 4 seconds. After one session of motor imagery (*t*=6), there was a 3-second rest period and a gray cross appeared on the screen. Each participant performed a total of 3 runs, each run consisting of 20 trials. Therefore, a total of 60 trials were performed (15 trials each for the left hand, right hand, feet, and tongue). After completing each run, the participants rested for 3 minutes to prevent the participants from generating too much noise data that would affect the training of the model due to fatigue or lack of concentration due to long-term repeated experiments[27]. The duration of the entire experiment with participants was approximately 15 minutes.

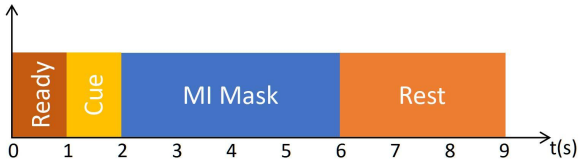

Figure 5   Data acquisition paradigm.

## 4.2. Preprocessing
### 4.2.1. Spatial Filtering

The electroencephalogram records the potential difference between each electrode and the reference electrode. However, each electrode can be influenced by noise that may be present in the reference electrode. To improve the signal-to-noise ratio, we employed the common average reference (CAR), where the new reference was the average of the electrical activity measured on all channels. The CAR filter removed common internal and external noise sources, leaving only exclusive activities of each EEG signal in each channel[28]. After applying CAR, the potential of each electrode can be calculated as follows[29]:

$$x_i^{CAR}(t) = x_i(t) - \frac{1}{C}\sum_{j=1}^{C} x_j(t) \quad (1)$$

Among them, $x_i^{CAR}(t)$ represents the spatial filtering output of the *i*-th electrode, $x_j(t)$ is the potential difference between the *j*-th electrode and the reference electrode, and "*C*" is the total number of electrodes, as illustrated in Figure 6 in this paper.

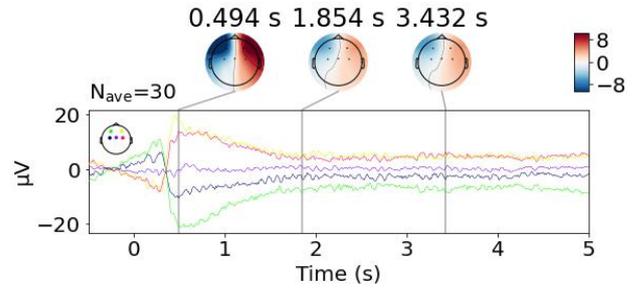

Figure 6   EEG data after spatial filtering.

### 4.2.2. Frequency Filtering

During the process of motor imagery, the primary motor cortex in the participants' brains records motor imagery signals, which exhibit changes in the Alpha and Beta rhythms. These rhythms include event-related desynchronization (ERD) and event-related synchronization (ERS), which can reflect how the brain responds to different tasks and stimuli[30]. ERD refers to a decrease in the amplitude of specific frequency bands in the EEG when a particular task or stimulus occurs, while ERS refers to an increase in the amplitude of specific frequency bands in the EEG during the same conditions. Both ERD and ERS reflect changes in the brain's spectral response to different tasks and stimuli, providing crucial information about brain function and neural network activity. To fully capture ERD and ERS, this paper used filters to extract the Alpha band and Beta band (with frequencies respectively in the ranges of 8–14Hz and 15–30Hz [31]) and applied an 8-30Hz filter for the filtering process.

### 4.2.3. Data Augmentation

Taking dataset 2a as an example, based on the dataset's description and Figure 2, it is evident that only seconds 2-6 are relevant to the motor imagery task in each experiment. Considering a sampling frequency of 250Hz, each experiment records 1000 valid samples. To increase the number of effective samples, a sliding window of 3.5 seconds was designed to slide over the data from seconds 2-6. The window slid in increments of 0.1 seconds, resulting in a total of five slides. Each segment obtained from these slides was assigned the same labels as the original experiment (increasing the number of experimental samples by fivefold).

## 5. Experimental Results
### 5.1. Baseline Models

The 3D-CLMI model was compared to several traditional machine learning (ML) and deep learning (DL) models (cross-entropy loss functions and Adam optimization algorithms were also used in these deep learning models):

(1) CSP-SVM[32]: Common spatial patterns (CSP) calculates covariance matrices between different brain states or motor intentions, selects a feature projection matrix, and maps the original features to a new feature space. In the projected feature space, the variance between different classes is maximized, making it easier to linearly separate different classes. CSP aims to enhance features in

EEG signals related to different brain states or motor intentions, thereby improving classification performance. Support vector machine (SVM), on the other hand, aims to build a classifier with good generalization ability by constructing an optimal hyperplane that separates samples from different classes, allowing for accurate classification of unknown samples.

(2) CSP-LCD[33]: Unlike CSP-SVM, which uses SVM for classification, LCD stands for linear discriminant classifier. LCD uses linear discriminant analysis (LDA) to classify the features extracted by CSP. LDA projects data into lower dimensions, aiming to keep the projection points of the same class as close as possible and those of different classes as far apart as possible.

(3) 2D CNN[34]: This model uses four convolutional layers and two pooling layers to extract spatial features from EEG data. Three dropout layers are inserted in between to prevent overfitting. Finally, fully connected layers with softmax activation are added for feature classification.

(4) 2D CNN-LSTM[19]: Building upon the 2D CNN approach, this model incorporates LSTM to extract temporal features from EEG data. The process involves first convolving EEG data and then passing the convolved data as input to LSTM for temporal feature extraction.

(5) EEGNet[35]: EEGNet consists of two main parts: depthwise separable convolution layers and spatio-temporal convolution layers. The depthwise separable convolution layers use two consecutive convolution operations, depthwise convolution, and pointwise convolution, to capture spatial and channel information in EEG data. The spatio-temporal convolution layers process the temporal dimension of EEG data using one-dimensional convolution, learning temporal patterns and sequence features within the EEG data.

(6) FBSF-TSCNN[36]: FBSF-TSCNN employs the filter bank spatial filtering (FBSF) algorithm to return EEG data with spatial filtering represented in time. Then, a time-space convolutional neural network (TSCNN) learns discriminative features from intermediate EEG data and provides decoding results.

(7) DeepConvNet[37]: DeepConvNet consists of a block with temporal convolution layers, a spatial convolution layer, and max-pooling layers, along with three blocks containing convolution layers and max-pooling layers.

## 5.2. Evaluation Metrics

To evaluated the performance of the classification model in this study, two evaluation metrics were used: accuracy and F1-score. Accuracy demonstrates the model's classification effectiveness, while F1-score, as the harmonic mean of precision and recall, provides a more accurate measure of model performance. The formulas are as follows[38]:

$$\text{Accuracy} = \frac{TP + TN}{TP + TN + FP + FN} \quad (2)$$

$$\text{Precision} = \frac{TP}{TP + FP} \quad (3)$$

$$\text{Recall} = \frac{TP}{TP + FN} \quad (4)$$

$$F1-\text{score} = \frac{2 \times \text{Precision} \times \text{Recall}}{\text{Precision} + \text{Recall}} \quad (5)$$

Among them, TP represents the number of true positives, FP represents the number of false positives, TN represents the number of true negatives, and FN represents the number of false negatives.

Given that BCI Competition IV 2a involves a four-class classification task, whereas the conventional F1-score is primarily used for binary classification tasks, this paper utilized the micro-average F1-score. The micro-average F1-score is computed based on overall statistics across all classes and considers the contribution of each sample. The calculation method involves summing the true positives, false positives, and false negatives across all classes and then computing the overall precision, recall, and F1-score[39].

During the training process, while ensuring the balanced distribution of samples, 80% of the entire dataset was randomly allocated to the training set, and the rest was allocated to the test set. To fine-tune individual model parameters effectively, minimize biases introduced by dataset selection randomness, and utilized the available data to its fullest extent, a 5-fold cross-validation strategy was employed[40]. Finally, for greater accuracy, the results from these 5 experiments were averaged. In the training of each deep learning network, 50 epochs were trained with a batch size of 64. To prevent overfitting, dropout layers were included in the model, and a learning rate decay strategy was applied[41]. Specifically, the learning rate was reduced by 30% every 10 epochs during training.

## 5.3. Experimental Results

### 5.3.1. Experimental Results on BCI Competition IV 2a Dataset

Table 2 presented the average classification accuracy results of the 3D-CLMI model proposed in this paper and other models on the public dataset. The 3D-CLMI model achieved the highest classification accuracy, surpassing other motor imagery classification models. The classification accuracy reached 92.7%, with a peak performance of 98.6%. Both the 2D CNN-LSTM and the 3D-CLMI model exhibited classification accuracies exceeding 85%, which was significantly better than other models. This indicates that the fusion of CNN and LSTM methods is more suitable for MI-EEG signal classification compared to machine learning methods (CSP-LCD, CSP-SVM) and pure CNN methods (2D CNN, EEGNet, FBSF-TSCNN, DeepConvNet), resulting in a substantial enhancement in classification accuracy. Among the nine participants, seven achieved the highest average classification accuracy under the 3D-CLMI model. Furthermore, statistical significance was observed in the classification accuracy between the proposed 3D-CLMI model and other models on this dataset through the Kruskal-Wallis

test[42]($p < 0.05$).

Table 2  Classification accuracy of 3D-CLMI on BCI Competition IV 2a.

|  | Ours | CSP-SVM | CSP-LCD | 2D CNN | 2D CNN-LSTM | EEGNet | FBSF-TSCNN | DeepConvNet |
|---|---|---|---|---|---|---|---|---|
| s01 | **0.948** | 0.776 | 0.724 | 0.865 | 0.941 | 0.844 | 0.810 | 0.844 |
| s02 | **0.872** | 0.479 | 0.583 | 0.622 | 0.733 | 0.501 | 0.468 | 0.531 |
| s03 | **0.979** | 0.774 | 0.821 | 0.898 | 0.933 | 0.881 | 0.838 | 0.868 |
| s04 | **0.898** | 0.547 | 0.523 | 0.656 | 0.877 | 0.631 | 0.523 | 0.625 |
| s05 | 0.864 | 0.546 | 0.625 | 0.511 | **0.865** | 0.657 | 0.315 | 0.403 |
| s06 | 0.878 | 0.394 | 0.614 | 0.484 | **0.886** | 0.602 | 0.426 | 0.569 |
| s07 | **0.938** | 0.833 | 0.829 | 0.860 | 0.898 | 0.866 | 0.773 | 0.868 |
| s08 | **0.986** | 0.808 | 0.885 | 0.784 | 0.889 | 0.799 | 0.755 | 0.778 |
| s09 | **0.983** | 0.581 | 0.834 | 0.760 | 0.866 | 0.827 | 0.736 | 0.806 |
| mean | **0.927** | 0.638 | 0.715 | 0.716 | 0.876 | 0.734 | 0.627 | 0.699 |
| std | 0.047 | 0.152 | 0.125 | 0.146 | 0.057 | 0.130 | 0.183 | 0.161 |
| *p*-value | / | <0.05 | <0.05 | <0.05 | <0.05 | <0.05 | <0.05 | <0.05 |

As shown in Figure 7, the 3D-CLMI model proposed in this paper exhibited the highest F1-score, with an average F1-score of 0.91, demonstrating its excellent classification capabilities for MI-EEG. Additionally, both the 3D-CLMI model and the 2D CNN-LSTM model achieved a high F1-score, with an average F1-score exceeding 0.8. This performance is significantly better than that of other models, indicating that the CNN-LSTM structure is well-suited for MI-EEG signal classification.

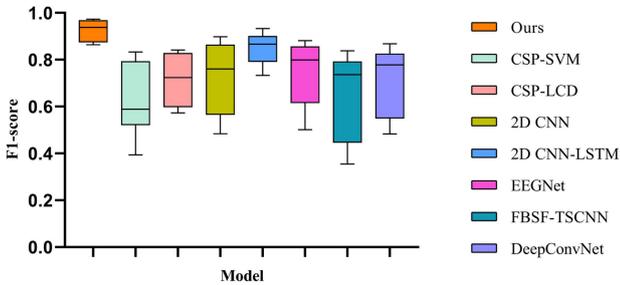

Figure 7  Average F1-score comparison of different models on BCI Competition IV 2a.

Furthermore, dimensionality reduction and visualization were performed using *t*-distributed stochastic neighbor embedding (*t*-SNE)[43]. As shown in Figure 8, the legends 0-3 represented the four class labels for left hand, right hand, feet, and tongue, distinguished by different colors. The left plot represented the distribution of the original signals, while the right plot represented the distribution after feature extraction by the 3D-CLMI model. It can be observed that the raw MI-EEG signals of different classes were challenging to distinguish. However, after feature extraction and classification by the model proposed in this paper, the distances between different class signals gradually increased, confirming its excellent performance in classification.

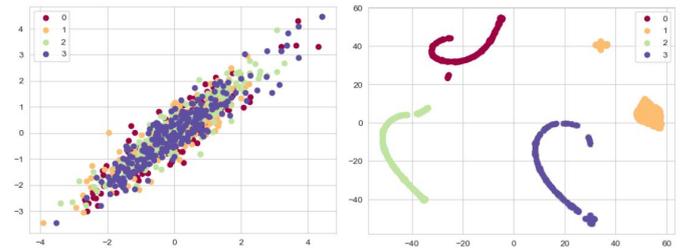

Figure 8  Two-dimensional representation of signal distribution on BCI Competition IV 2a.

### 5.3.2. Experimental Results on Our Dataset

Table 3 presented the average classification accuracy using the 3D-CLMI model and other models on our dataset. Similar conclusions to those in Section 5.3.1 can be drawn: (1) The 3D-CLMI model achieved the highest classification accuracy, significantly outperforming other models. (2) The CNN-LSTM fusion model was well-suited for the classification of motor imagery. (3) Among the 12 participants, 11 of them had the highest average classification accuracy when using the 3D-CLMI model for classification. Additionally, Kruskal-Wallis tests indicated a significant difference in classification accuracy between the 3D-CLMI model and the other models ($p < 0.05$).

Table 3    Classification accuracy of 3D-CLMI on our dataset.

|  | Ours | CSP-SVM | CSP-LCD | 2D CNN | 2D CNN-LSTM | EEGNet | FBSF-TSCNN | DeepConvNet |
|---|---|---|---|---|---|---|---|---|
| s01 | **0.956** | 0.825 | 0.863 | 0.914 | 0.942 | 0.921 | 0.894 | 0.843 |
| s02 | **0.917** | 0.673 | 0.841 | 0.875 | 0.827 | 0.887 | 0.867 | 0.895 |
| s03 | **0.924** | 0.826 | 0.868 | 0.846 | 0.915 | 0.846 | 0.879 | 0.837 |
| s04 | **0.979** | 0.854 | 0.866 | 0.937 | 0.903 | 0.635 | 0.916 | 0.889 |
| s05 | **0.988** | 0.617 | 0.899 | 0.827 | 0.954 | 0.837 | 0.903 | 0.939 |
| s06 | **0.935** | 0.878 | 0.817 | 0.929 | 0.899 | 0.793 | 0.831 | 0.846 |
| s07 | **0.963** | 0.842 | 0.904 | 0.817 | 0.926 | 0.875 | 0.815 | 0.773 |
| s08 | **0.984** | 0.884 | 0.736 | 0.964 | 0.937 | 0.946 | 0.887 | 0.836 |
| s09 | **0.973** | 0.775 | 0.827 | 0.927 | 0.871 | 0.864 | 0.724 | 0.791 |
| s10 | **0.962** | 0.896 | 0.875 | 0.736 | 0.912 | 0.758 | 0.937 | 0.806 |
| s11 | 0.938 | 0.793 | 0.684 | 0.870 | **0.943** | 0.779 | 0.930 | 0.842 |
| s12 | **0.951** | 0.814 | 0.838 | 0.846 | 0.879 | 0.862 | 0.877 | 0.859 |
| mean | **0.956** | 0.806 | 0.835 | 0.874 | 0.909 | 0.834 | 0.872 | 0.846 |
| std | 0.025 | 0.081 | 0.062 | 0.062 | 0.035 | 0.080 | 0.056 | 0.044 |
| $p$-value | / | <0.05 | <0.05 | <0.05 | <0.05 | <0.05 | <0.05 | <0.05 |

Similarly, the 3D-CLMI model and the 2D CNN-LSTM model exhibited a significantly higher average F1-score compared to other models, with the 3D-CLMI model achieving the highest F1-score, reaching an average of 0.95. This indicated that the CNN-LSTM structure is well-suited for the classification of MI-EEG signals. Furthermore, all models showed improved F1-score compared to dataset 2a (all F1-scores above 0.6), and most models had reduced F1-score variances. This suggested that the MI-EEG signal sampling quality on our dataset was higher, resulting in more stable classification performance.

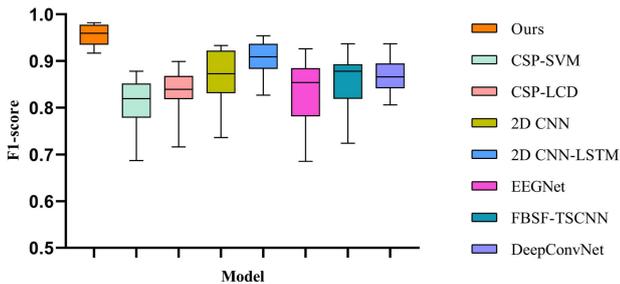

Figure 9    Average F1-score comparison of different models on our dataset.

In Figure 10, the legends 0-3 represented the four classes: left hand, right hand, feet, and tongue, respectively, distinguished by different colors. Similarly to dataset 2a, there was an initial challenge in distinguishing the raw MI-EEG signals on our dataset. However, after feature extraction and classification by the 3D-CLMI model, the distances between signals of different labels increased, making them easier to distinguish.

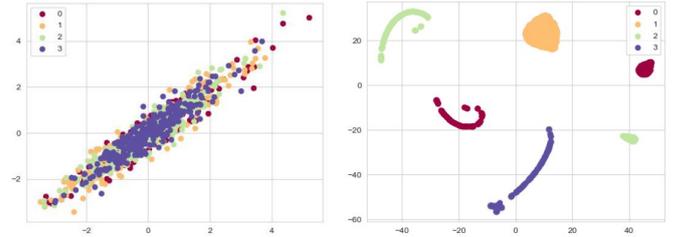

Figure 10    Two-dimensional representation of signal distribution on our dataset.

### 5.4. Ablation Experiment

As shown in Table 4, on BCI Competition IV 2a, ablative experiments were conducted on different modules within 3D-CLMI to analyze the contributions and importance of each module[44]. The primary focus was on two main parts of 3D-CLMI: (1) The convolution method, which referred to the dimensions of the used convolutional kernels. (2) The connection mode between CNN and LSTM, which can be either serial or parallel.

Table 4    Ablation experimental results.

| Convolution method | Network structure | Average accuracy | $p$-value |
|---|---|---|---|
| 2D CNN | CNN-LSTM parallel | 88.3% | <0.05 |
| 3D CNN | CNN-LSTM serial | 90.9% | <0.05 |
| 3D CNN | CNN-LSTM parallel | **92.7%** | / |

According to Table 4, it can be observed that using 3D convolutional kernels improved accuracy by 4.4%, and changing from a serial network structure to a parallel network structure increased accuracy by 1.8%. It was evident that 3D convolutional kernels play a crucial role in feature extraction, and the CNN-LSTM parallel model structure used in this paper is also helpful in improving classification accuracy. Furthermore, through Wilcoxon signed-rank tests conducted between the 3D CNN-LSTM parallel structure model and the other two ablated models, it was found that there were significant differences between these models ($p<0.05$). This also indicated that the proposed model in this paper effectively enhanced network performance.

**5.5. Parameter Sensitivity Test**

To explore the impact of network parameters on the model in this paper, three important parameters were selected: learning rate, the number of LSTM units, and training epochs, and differences between these parameter influences were discussed.

The results in Table 5 indicated that the average classification accuracy varies slightly with changes in the learning rate. The 3D-CLMI model achieved the highest classification accuracy when the learning rate was set to 0.001. It can be concluded that choosing a small learning rate or a large learning rate did not lead to good accuracy. This is because a small learning rate (*e.g.*, 0.00001) results in an extremely slow convergence process, while a large learning rate (*e.g.*, 0.1) can lead to non-convergence or even divergence.

Table 5   Impact of learning rate on 3D-CLMI.

| Learning rate | Average classification accuracy |
|---|---|
| 0.1 | 0.896 |
| 0.01 | 0.904 |
| 0.001(Ours) | **0.927** |
| 0.0001 | 0.912 |
| 0.00001 | 0.898 |

LSTM units are the basic building blocks of LSTM networks, used for modeling and remembering sequential data. The number of LSTM units determines the classification performance of the LSTM network[45]. From the results in Table 6, it can be observed that when the number of LSTM units was set to 256, the 3D-CLMI model achieved the highest classification accuracy. As the number of LSTM units decreased (*e.g.*, from 256 to 64), the average classification accuracy significantly decreased. However, increasing the number of LSTM units (*e.g.*, from 256 to 1024) did not lead to an improvement in classification accuracy, instead, it slightly decreased.

Table 6   Impact of LSTM units on 3D-CLMI.

| Number of LSTM units | Average classification accuracy |
|---|---|
| 64 | 0.824 |
| 128 | 0.866 |
| 256(Ours) | **0.927** |
| 512 | 0.922 |
| 1024 | 0.916 |

The number of training epochs determines the model's ability to learn from the training data. Appropriate training epochs give the model more opportunities to learn complex patterns and relationships from the data, improving its ability to fit the training data[46]. From the results in Table 7, it can be seen that when the number of training epochs was set to 100, the 3D-CLMI model achieved the highest classification accuracy of 92.7%. As the number of training epochs increased, the accuracy gradually improved from 86.5% to 92.7%. However, once a sufficient number of epochs for training was reached, the accuracy remained within a certain range (approximately 92.5%). However, too many training rounds may cause the model to overfit, so it is not recommended.

Table 7   Impact of epochs on 3D-CLMI.

| Number of epoch | Average classification accuracy |
|---|---|
| 30 | 0.865 |
| 50 | 0.923 |
| 100(Ours) | **0.927** |
| 150 | 0.924 |
| 200 | 0.925 |

## 6. Discussion

In this study, the performance of the 3D-CLMI model in processing MI-EEG signals was fully assessed. To conclude, the main innovation points of the 3D-CLMI model are as follows: First, the strong correlation between different EEG channels leads to strong spatial features between different EEG channels. Based on this characteristic, superimposing the two-dimensional MI-EEG data of multiple channels into three-dimensional data and using a 3D convolution kernel to extract the features of multiple channels of EEG signals is more efficient than using a 2D convolution kernel to extract the two-dimensional features of each channel individually. Second, the convolution module in 3D-CLMI uses multiple convolution kernels of different sizes for convolution and combines the features extracted by different convolution kernels. For the feature extraction of MI-EEG signals at this stage, it is impossible to determine what scale of convolution kernel to use to extract more appropriate features, so multiple convolution kernels of different scales are used to extract features. Since multi-kernel convolution superimposes receptive fields of different sizes, features of different granularities can be better extracted through multi-kernel convolution[47]. Finally, the reason why the parallel structure of CNN and LSTM is used is that the currently commonly used CNN-LSTM serial structure may cause the temporal features of the raw data to disappear to a certain extent during the convolution process, resulting in the convolution of the data. When inputting LSTM, the effect of temporal feature extraction is not ideal. Using the CNN-LSTM parallel structure, the spatial and temporal features of the raw data can be extracted separately, concatenated, and then classified together, so that the spatial and temporal features of the raw MI-EEG data can be retained to the greatest extent.

However, there were still some shortcomings in this paper, the experiment in this paper was based on a single-participant training model and classifying motor imagery, but the classification effect will decrease under the conditions of cross-participant experiments. In the next step, methods such as transfer learning with pre-trained models will be considered to enhance the classification effect of the classification model under cross-participant conditions. In addition, the current experiment was an offline experiment. In the future, this model will be expanded to be applied to more classification tasks, and the EEG signal will be dynamically segmented to achieve real-time analysis of user intentions so that it can be applied to autonomous driving and medical rehabilitation[48] and other emerging fields.

## 7. Conclusion

This paper introduced the 3D-CLMI model for the classification of MI-EEG signals and provided an explanation and analysis of the key modules and principles of our proposed model. Then conducted performance evaluation experiments with other commonly used MI-EEG signal classification models on public dataset BCI Competition IV 2a and our dataset. The results demonstrated that on the BCI Competition IV 2a dataset, our model achieved an average classification accuracy of 92.7% (with a peak of 98.6%) and an F1-score of 0.91. On our four-class MI experiment dataset, our model obtained an average classification accuracy of 95.6% (with a peak of 98.8%) and an F1-score of 0.95. These results clearly surpassed the performance of existing classification models.

## Acknowledgments

The authors would like to thank all the volunteers who participated in the experiments. This work was supported in part by the Natural Science Foundation of Zhejiang Province under Grant LR22F020003, and the National Natural Science Foundation of China under Grant 62172368.